\def\year{2021}\relax
\documentclass[letterpaper]{article} 
\usepackage{aaai21}  
\usepackage{times}  
\usepackage{helvet} 
\usepackage{courier}  
\usepackage[hyphens]{url}  
\usepackage{graphicx} 
\usepackage{amsmath}
\usepackage{amssymb}
\usepackage{comment}
\usepackage{natbib}  
\usepackage{caption} 
\title{Hybrid Information-driven Multi-agent Reinforcement Learning}
\author{
    William A. Dawson,
    Ruben Glatt,
    Edward Rusu,
    Braden C. Soper,
    Ryan A. Goldhahn
    \\
}
\affiliations{
    Lawrence Livermore National Laboratory\\
    7000 East Ave\\
    Livermore, California  94550\\
    \{dawson29, glatt1, rusu1, soper3, goldhahn1\}@llnl.gov
}

\begin{document}

\maketitle

\begin{abstract}
Information theoretic sensor management approaches are an ideal solution to state estimation problems when considering the optimal control of multi-agent systems, however they are too computationally intensive for large state spaces, especially when considering the limited computational resources typical of large-scale distributed multi-agent systems.
Reinforcement learning (RL) is a promising alternative which can find approximate solutions to distributed optimal control problems that take into account the resource constraints inherent in many systems of distributed agents.
However, the RL training can be prohibitively inefficient, especially in low-information environments where agents receive little to no feedback in large portions of the state space.
We propose a hybrid information-driven multi-agent reinforcement learning (MARL) approach that utilizes information theoretic models as heuristics to help the agents navigate large sparse state spaces, coupled with information based rewards in an RL framework to learn higher-level policies.
This paper presents our ongoing work towards this objective.
Our preliminary findings show that such an approach can result in a system of agents that are approximately three orders of magnitude more efficient at exploring a sparse state space than naive baseline metrics.
While the work is still in its early stages, it provides a promising direction for future research.
\end{abstract}

\section{Introduction}
The continuing technical advancement of sensors together with the reduction in their size and production costs allow for increasingly complex sensor applications driven by machine learning and artificial intelligence algorithms.
An observable trend is the integration of heterogeneous, autonomous sensors into complex, intelligent sensor networks that can be controlled in a centralized \cite{chen2019iraf} or decentralized manner \cite{wang2020federated}.
However, these advancements also come with a number of open challenges associated with the design and control of such networks.
Many of those challenges are related to scaling the algorithmic complexity as the number of agents increases.
In particular, many exact centralized methods for learning optimal control policies become prohibitively expensive as the number of sensors grows, suffering from the so-called ``curse of dimensionality'' \cite{Boutilier1996mmdp}.
Such computational challenges are compounded when sensors and actors are combined in intelligent, autonomous agents that can work together to solve highly complex tasks \cite{baker2019emergent}.
Moreover, when considering large-scale decentralized networks composed of low-to-mid cost sensors, resource constraints must be taken into account to ensure the sensor network accomplishes its objective given a limited resource budget (e.g., communication bandwidth, battery life, CPU/GPU cycles, reach, etc.).

One basic way to describe such systems in which decision-making agents interact to solve a task is the extension of Markov Decision Processes (MDP) \cite{puterman2014markov} to the multi-agent case, a Markov game or simply Multi-agent MDP (MMDP) \cite{littman1994markov}.
An MMDP is defined by a set of states $S$, a set of actions $A_{i}$ for each agent $i={1,...,n}$ in the environment, a transition function $T(s^{'}|s,u)$ defining the probability of observing a follow-up state $s^{'}$ after applying the joint action of all agents $u$ in state $s$, and a reward function $R(s,u)$ providing a reward for applying the joint action $u$ in state $s$.
Here, we consider stochastic games which are a further variant where the reward function $R_{i}(s,u)$ provides individual rewards for each agent instead of a joint reward \cite{bowling2000analysis}.
Many possible ways to solve this kind of decision-making processes are described in the field of Reinforcement Learning (RL) \cite{sutton2018reinforcement} and Multi-agent Reinforcement Learning (MARL) \cite{busoniu2008comprehensive}.
In particular, single-agent learning has received much attention and the literature provides solutions on a wide range of challenging tasks such as Atari game playing \cite{mnih2015human} or electric vehicle charging \cite{pettit2019increasing}.
In general, RL has been shown to find good solutions for tasks involving distributed systems, but scalability remains an issue due to the high sample inefficiency and long training time required in complex systems, as evidenced, for example, by training over 900 agents each for 44 days (!) to find a good solution in the real-time strategy game StarCraft II \cite{vinyals2019grandmaster}.

Most works consider agents that try to learn very complex problems without making many assumptions or using prior knowledge, for example known physical processes in an environment. While minimizing assumptions and prior knowledge may lead to more robust policies, it nevertheless makes the learning process extremely difficult in complex problems.
In this work, we investigate agents that use model heuristics and approximations when available and reasonable, so that the RL agent can focus on learning the aspects of the problem for which we do not have reasonable models.
Specifically, we consider agents trying to predict the source location of a chemical plume in a partially observable environment. The agents make plume concentration measurements and then decide how to explore the state space based on information theoretic metrics.
Our experiments show that by combining these approaches, along with communication between the agents, we can achieve a speedup in learning while reducing computational requirements.
This report presents the general idea of a hybrid information-based RL approach as well as some preliminary experiments and results from our ongoing work.
We also discuss the important challenges that remain with such an approach.

\section{The Plume State Estimation Problem}
This work focuses on a multi-agent chemical plume source localization problem, where multiple agents are allowed to move around a two-dimensional space taking concentration measurements in concert to estimate the location of the chemical plume's source \cite[see][for more details]{schmidt2019sequential}.
Our initial attempts at a pure RL solution to this problem appeared to be computationally intractable as the state space was relatively large, facing the typical challenges of multi-agent learning \cite[e.g.,][]{vinyals2019grandmaster} plus the additional challenge that the vast majority of the agents' initial measurements were null (i.e., the plume occupies a very small region of the environment).
This motivated us to consider a new hybrid information-driven multi-agent RL approach.

\section{Information-based Motion and Exploration}
Arguably the most fundamental solution to state estimation problems is the information theoretic sensor management approach \cite{hero2008infotheory}.
Following \cite{hero2007foundations}, we present the relevant details of Information-Optimal Policy Search which will become the core of our hybrid RL approach.

The basic principle is that at time $t$ we want to explore a set of possible locations at which we will make future measurements. Let  $m_{t}$ be the observation made at time $t$ and let $(x_{t},y_{t})$ be the location where the measurement was taken at time $t$. We denote the true fixed source location by $(x_{\text{s}},y_{\text{s}})$. The observations are assumed to be conditionally Gaussian given the location of the observation, the true source location and a parametric plume model given by a deterministic function $f$, giving us $m_t|x_t, y_t, x_{\text{s}}, y_{\text{s}} \sim \mathcal{N}(f(x_t, y_t, x_{\text{s}}, y_{\text{s}}), \sigma^2)$. Details on the function $f$ can be found in \cite{schmidt2019sequential}. We assume each agent places a prior distribution $\pi(x_{\text{s}},y_{\text{s}})$ over the source location parameters.  
We assume a discretized space of possible measurement and source locations with $x_t \in A$, $y_t \in B$, $x_{\text{s}} \in I$, and $y_{\text{s}} \in J$
(discrete space is not a requirement of the method in general, although it facilitates an FFT approximation).

The objective is to choose the next measurement location $(x_{t+1}, y_{t+1})$ that maximizes the expected information gain (IG) relative to the cost $C(x_{t+1}, y_{t+1})$ of making that measurement. The IG is defined as the $\alpha$-divergence \cite{renyi1961alpadiv} between a prior distribution on the source location and the posterior distribution of the source location given a new measurement $m_{t+1}$ taken at location $(x_{t+1}, y_{t+1})$. In the limit that $\alpha$ approaches 1 this reduces to the Kullback-Leibler divergence $D_{KL}$. 

Letting $d_t = (x_t, y_t, m_t)$ be the data collected at time $t$, and $d_{1:t} = (x_{1:t}, y_{1:t}, m_{1:t})$ be the data collected up until time $t$, the information gain is defined as

\begin{multline*}
  \mathrm{IG}(m_{t+1}| x_{t+1}, y_{t+1}) \equiv D_{\mathrm{KL}}\left(p(x_{\text{s}}, y_{\text{s}}| d_{1:t+1}) || q(x_{\text{s}},y_{\text{s}})\right)\\
  =  \sum_{x_{\text{s}}} \sum_{y_{\text{s}}} p(x_{\text{s}}, y_{\text{s}}| d_{1:t+1})  \log \frac{p(x_{\text{s}}, y_{\text{s}}| d_{1:t+1})}{q(x_{\text{s}},y_{\text{s}})}.
    \label{equ:ig}
\end{multline*}
The term $q(x_{\text{s}}, y_{\text{s}})$ in the above equation could take on various probability distributions. However, in the multi-agent RL case it is important that metrics be consistent across agents and epochs. 
Thus, we set this as the initial prior $\pi(x_{\text{s}}, y_{\text{s}})$.

Because we do not know the true source location $(x_{\text{s}}, y_{\text{s}})$, we must marginalize over the posterior predictive distribution of the observation $m_{t+1}$ in order to determine the expected IG. Letting 
\begin{multline*}
p(m_{t+1}|x_{t+1}, y_{t+1}, d_{1:t}) =  \\ \sum_{x_{\text{s}}} \sum_{y_{\text{s}}}  p(m_{t+1}| x_{t+1}, y_{t+1}, x_{\text{s}}, y_{\text{s}})p(x_{\text{s}},y_{\text{s}} | d_{1:t})
\end{multline*}
be the posterior predictive distribution of the measurement $m_{t+1}$ taken at location $x_{t+1}, y_{t+1}$ given previously observed data $d_{1:t}$, we have the expected IG as 
\begin{multline*}
  \mathbb{E}_{m_{t+1}}\left[\mathrm{IG}(m_{t+1}| x_{t+1}, y_{t+1})\right]=\\
  \int_{m_{t+1}}\mathrm{IG}(m_{t+1}| x_{t+1}, y_{t+1}) 
  p(m_{t+1}|x_{t+1}, y_{t+1}, d_{1:t}) 
  dm_{t+1}.
\end{multline*}
The key implication of the previous equation is that it results in another set of sums outside of the IG sums, since we are implicitly determining the IG for the expected measurement at a possible combination of source and measurement locations,
\begin{multline*}
  \mathbb{E}\left[m_{t+1}(x_{t+1}, y_{t+1}, x_s, y_s)\right] = \\
  \sum_{x_{\text{s}}} \sum_{y_{\text{s}}}  f(x_{t+1}, y_{t+1}, x_{\text{s}}, y_{\text{s}})p(x_{\text{s}},y_{\text{s}} | d_{1:t}).
\end{multline*}

Note that these expected IG must be computed for each possible measurement location $(x_{t+1},y_{t+1}) \in A \times B$. The objective can then be written in the following way. 
\begin{equation}
  \label{equ:ig_cost}
  \max_{\substack{x_{t+1} \in A, \\ y_{t+1} \in B}}  \frac{\mathbb{E}_{m_{t+1}}\left[\mathrm{IG}\left( m_{t+1} |  x_{t+1}, y_{t+1} \right)\right]}{C( x_{m_{t+1}},  y_{m_{t+1}})}
\end{equation}

The full information theoretic sensor management solution to this simple plume problem requires $A \times B \times I^{2} \times J^{2}$ total calculations per time step per agent.
Clearly as the state estimation space increases in size or dimensionality, the number of agents increases, or a non-myopic optimization is attempted, the problem becomes computationally intractable.
\citet{kreucher2008particle} suggest particle filtering as a possible solution, but in the case of a large state space with a uninformative prior, the number of particles necessary is initially approximately $A \times B \times I^{2} \times J^{2}$.
Thompson sampling is another alternative and can reduce the dimensionality of the problem to $A \times B \times I \times J$ and has proven convergence properties \cite{verstraeten2020thompson}.
We can further reduce the computational complexity to $I \times J \times \log(A \times B)$ by adopting a Signal to Noise Ratio (SNR) expectation maximization approximation \cite{fisher1935design} to IG expectation maximization and leverage the fact that the mathematical form of such an approximation facilitates fast Fourier transformation convolution.
Note that this last approximation breaks down when the signal is not weak. However, if this is the case then the problem is much simpler.

For this work we adopt the SNR expectation maximization approximation, as it is most conducive to embedding in RL algorithms and on-agent computation.
To demonstrate the information-based motion heuristic we randomly populated a space with five agents and a single plume as in \citet{schmidt2019sequential}.
The area of the plume where the agents could make an SNR$>$1 measurement is $\sim$0.003 times the total area covered by the uniform prior for the plume source location.
At each time step the agents take a local measurement, communicate their measurement to other agents, update their posterior distribution, and use the heuristic to decide where to move next.

\begin{figure}
  \centering
  \includegraphics[width=3.05in]{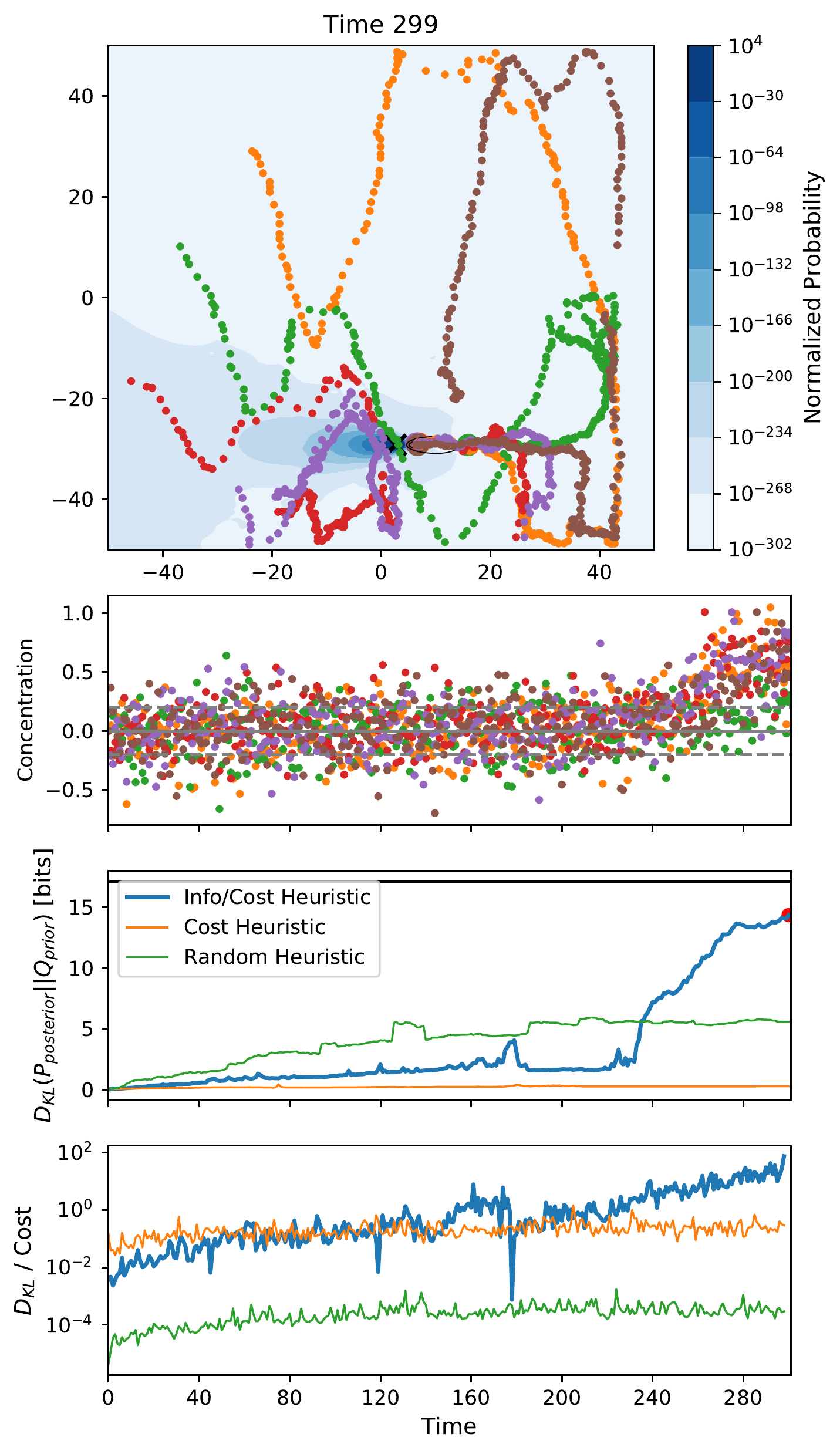}
  \caption{
  Results of a simplified five agent plume localization problem relying exclusively on an information-based motion heuristic.
  Top panel: The two-dimensional plume environment.
  This history of the agent measurement locations are shown with the small circles, colored according to agent ID.
  The contour background shows the normalized posterior probability distribution of the plume source location as estimated by at agents at time step 299, which has been concentrated about the true plume source location (black X).
  The thin black contours to the right of the plume source denote the region where the agents could make a SNR $> 1$ concentration measurement.
  Second panel: A history of the concentration measurements made by the agents, which for the most part are scattered about zero with variance due to user defined noise (dashed black lines denote $1\sigma$).
  Bottom two panels: The history of the information gain ($D_\mathrm{KL}$) for the simulation (thick blue line).
  For comparison purposes we show histories for two separate simulations using just the cost weight as a motion heuristic (orange) and completely random motion (green).
  Fluctuations are a byproduct of the noisy measurements.
  The information-based motion heuristic is $> 3$ orders of magnitude more efficient than the baseline random methods in the long run.
  }
  \label{fig:info_gain}
\end{figure}

In this simulation there is also a distance squared motion cost plus overhead associated with the movement options (Eq. 1), which is a physical model of the energy cost of accelerating the agent to a new location.
Fig.~\ref{fig:info_gain} shows a history of one of these simulations where the agents systematically explore the space, for the most part making null detections and ruling out possible plume source locations, until one agent makes a significant plume concentration measurement around time step 230.
At that point the plume source posterior becomes concentrated and the other agents naturally converge on the plume location, seamlessly transitioning from exploration to exploitation.
Note, however, that the maximum possible IG for this simulation is $\sim$17 bits but the agents only achieve IG $\sim$15 bits.
This is because the SNR expectation maximization approximation of IG expectation maximization breaks down in the high SNR regime.
Still, the plume source 95\% credible interval has been localized to just a few discrete locations.
As can be see from the lower panels of Fig.~\ref{fig:info_gain} the information-based motion heuristic is more than three-orders of magnitude more efficient than baseline random heuristics.
Perhaps just as important is that this approach based on Eq.~\ref{equ:ig_cost} enables the agents to naturally transition from exploration to exploitation.

\section{Hybrid Information-driven Multi-agent Reinforcement Learning}

As demonstrated above, information theory based sensor optimization is an intuitive, effective, and relatively computationally efficient multi-agent motion heuristic.
However, once we increase the complexity of the problem by no-longer allowing the agents to communicate, infer their state, and make measurements for free, but instead factor the costs of these choices into the optimization problem, the optimization cost quickly becomes prohibitive, and applying a fully information theory based sensor optimization approach becomes computationally intractable.
RL holds the potential to learn a good policy for this scenario, but as we previously noted the training was prohibitively computationally expensive for the full partially observable state space estimation with a noninformative prior.
Thus we propose an approach leveraging models where they exist as heuristics and only using RL for the high level decision making problem.
In our case, this means using RL to learn policies for decisions about what higher level action the agent should take and then leveraging the information based heuristic to determine the agent's motion.
Since both the heuristic and RL reward are based on the same IG/cost function this allows for consistent behavior and enables the possibility of easily transferring the agents decision making abilities to different tasks, which has been shown to be difficult in traditional approaches \cite{glatt2016towards}.

\begin{figure}
  \centering
  \includegraphics[width=2.in]{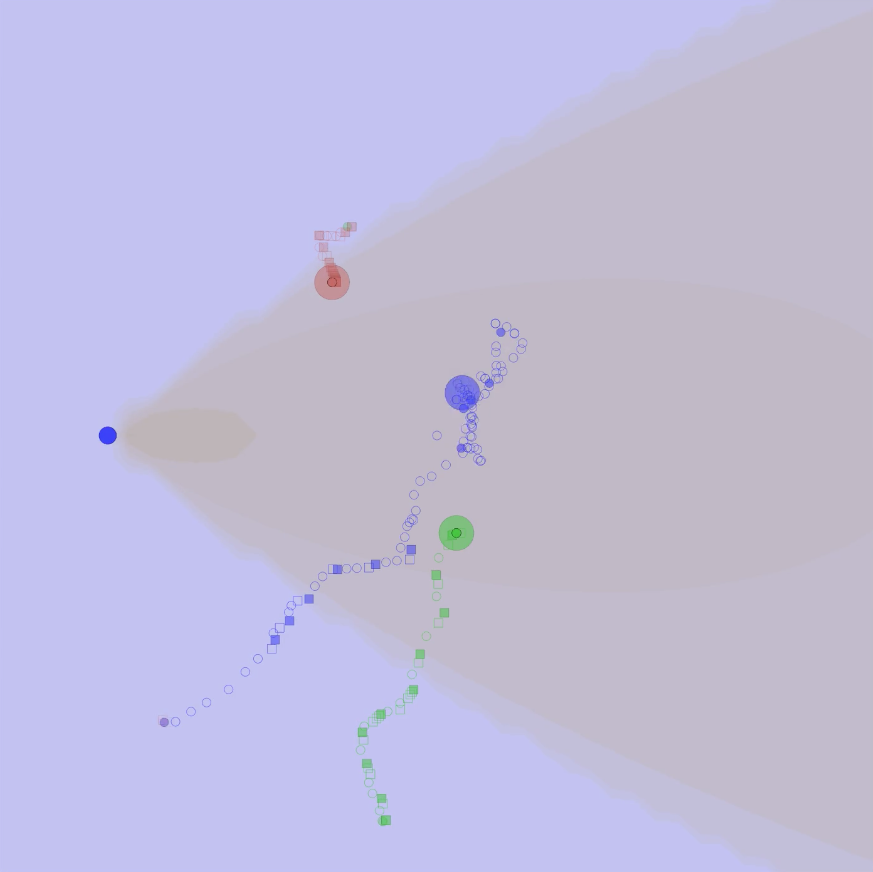}
  \caption{Gaussian plume model with plume source (small full blue dot) and concentration indication (orange) together with three agents with action trail (red, light blue, green) inter-acting in the environment.}
  \label{fig:environment}
\end{figure}

For our hybrid RL experiments, our environment considers a single plume source location that emits a static Gaussian plume as in the previous section (see Fig.~\ref{fig:environment}) and provides the physics of the world so that agents can move, interact with other objects (when present), and communicate with each other.
The plume concentration is normalized in the possible concentration range to keep all state variables in a similar range and to avoid issues when learning.

\textbf{Action space:} The action space for the agents is a set of five discrete actions: do nothing, move, take a measurement, update, and communicate.
\textit{Communicate} collects the last measurements from the other agents and directly updates the posterior, \textit{measuring} means taking a new concentration measure at the current location and saving it in the concentration buffer, and \textit{update} uses the last maximum four measurements after the last update to update the posterior and resets the concentration buffer.
The \textit{move} action then uses the agent's current belief (i.e., posterior probability distribution) about the location of the source to change the acceleration of the agent towards the estimated source location.

\textbf{Observation space:} The observation space is a flattened vector containing current agent position (x, y), agent velocity (x, y), wind velocity (x, y), last concentration measurement, current source location estimate (x, y), information gain since episode start, and some internal state indicators as Boolean values indicating if the agent moved since taking the last measurement, if the agent has performed the same action more than 4 times, and a one-hot vector indicating the last action taken (5 values here).

\textbf{Reward structure:} Our reward structure is driven by the internal assumptions of the agent using the distance between the actual source location and the agents' current best estimate of the source location as well as the information gain achieved so far.
However, we also integrated part of the reward based on logical assumptions that we make about favorable behavior with respect to cost of operation to reduce energy consumption and low profile operation to protect against detection.
In these initial settings, at the beginning of training, the information-based part is very dominant; but over time when the estimation gets better and the achieved information gain increases, this additional action based reward becomes more dominant.
The goal is to facilitate learning of desired behavior under logical assumptions without giving too much guidance, for example, when an agent has high confidence about the source location estimate it is best to \textit{do nothing} to save resources and avoid detection through movement or communication.

\begin{figure}
  \centering
  \includegraphics[width=3.3in]{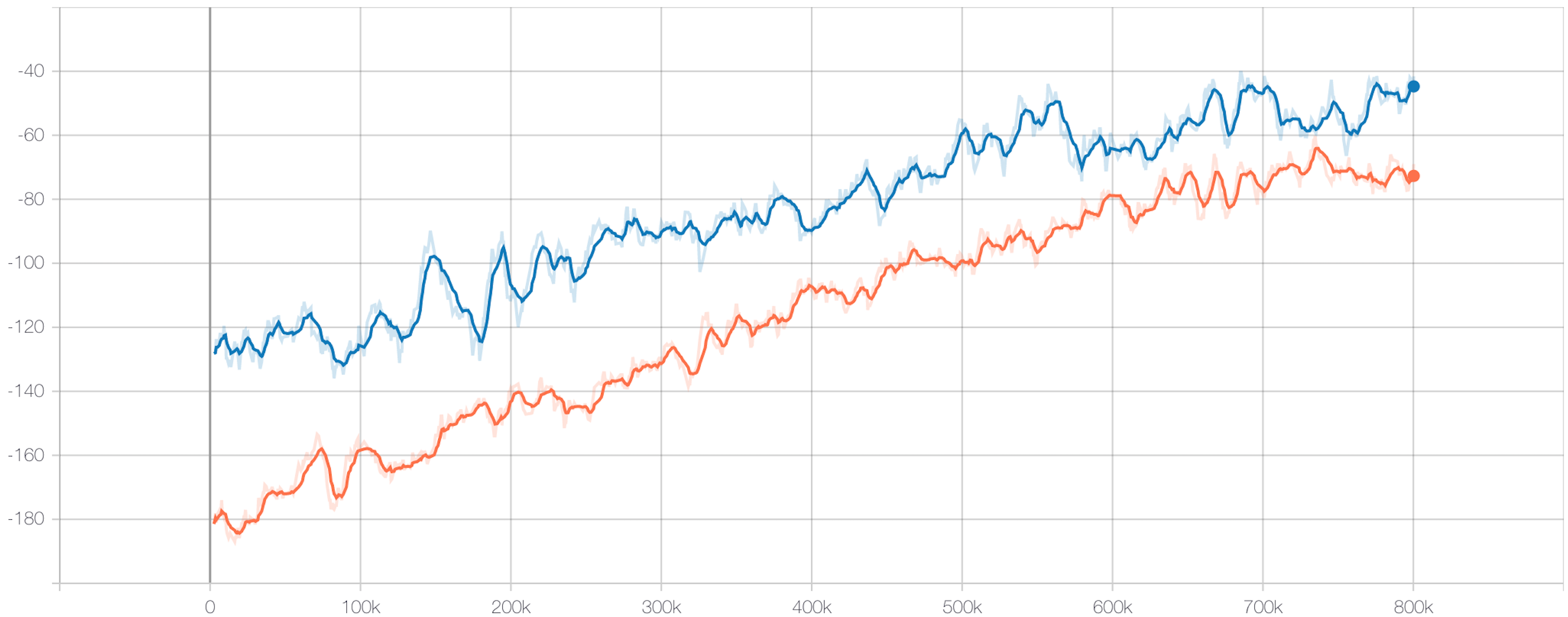}
  \caption{Reward over steps for agents that learn individually (red) and for agents that communicate with each other (blue).}
  \label{fig:MARL}
\end{figure}

\textbf{Results:} In the experiments, we used a small variant of the \textit{Deep Q-Network} \cite{mnih2015human} to learn the high-level actions of each agent while we use the information based estimation model for the location estimation and actual movement decision.
Our first experiments show that agents that communicate have a clear head start when it comes to estimating the source location properly and then remain ahead even when individual agents have already converged to a behavior (Fig.~\ref{fig:MARL}).
This is based on the fact that even bad measurements provide information about possible source locations and we observed that agents quickly learn to use the communication action early to get higher rewards from the start.

\section{Discussion}

Other approaches that use communication to speed up learning are often built on a student-teacher relationship where the teacher is usually an expert that advises the student with perfect information.
Agents that advise each other with imperfect information are also considered in \cite{da2017simultaneously}, where the authors show that confidence based communication can be a helpful tool to speed up learning.
While this approach investigates ad hoc communication with individual agents learning in the same environment, the approach only considers a pure RL approach without integrating additional intelligence.

An idea that could integrate well with our approach to improve resource conservation is \cite{da2020uncertainty} where agents ask for advice only when their epistemic uncertainty is high for a certain state.
An important aspect in this work is also that the proposed method considers that the advice is limited and might be sub-optimal.

\section{Conclusion \& Future Work}

In this short position paper we have shared our initial considerations with respect to hybrid information-driven multi-agent reinforcement learning.
State space estimation based on information metrics is a powerful tool that we can leverage to solve tasks where we use RL to learn a decision-making policy that indicates what to do while basic knowledge about physical processes guides us on how to do it.

In our experiments, agents cooperate only by sharing knowledge through communication which is triggered by the communication action.
In future versions, we intend to investigate other means of training the agents like Multi-Agent Deep Deterministic Policy Gradient (MADDPG) \cite{lowe2017multi} or Reinforced Inter-Agent Learning (RIAL) and Differentiable Inter-Agent Learning (DIAL) \cite{foerster2016learning}.
On the information side, we are currently working on integrating decentralized Markov chain Monte Carlo (MCMC) methods for full posterior inference for non-discretized, non-conjugate models.
With respect to scalability, we are working towards high-fidelity, discrete-event simulations to model wireless communication protocols and a larger number of agents.

\section{Acknowledgments}
This work was performed under the auspices of the U.S. Department of Energy by Lawrence Livermore National Laboratory under contract DE-AC52-07NA27344.
Lawrence Livermore National Security, LLC through the support of LDRD 20-SI-005.
LLNL-CONF-816423.

\bibliography{main}
\bibstyle{aaai21}

\end{document}